\documentclass[sigconf]{acmart}
\AtBeginDocument{%
  }

\setcopyright{acmlicensed}
\copyrightyear{2026}
\acmYear{2026}
\acmDOI{XXXXXXX.XXXXXXX}
\acmConference[Conference acronym 'XX]{Make sure to enter the correct
  conference title from your rights confirmation email}{June 03--05,
  2018}{Woodstock, NY}
\acmISBN{978-1-4503-XXXX-X/2018/06}

\usepackage{multirow}
\usepackage{array}
\usepackage{adjustbox} 
\usepackage{graphicx}   
\usepackage{subcaption} 
\usepackage{enumitem}
\begin{document}

\title{DAIAN: Deep Adaptive Intent-Aware Network for CTR Prediction in Trigger-Induced Recommendation}

\author{Zhihao Lv}
\affiliation{%
  \institution{Xianyu of Alibaba}
  \city{Hangzhou}
  \country{China}
}
\email{fengguang.lzh@taobao.com}
\authornote{All these authors contributed equally to this research.}

\author{Longtao Zhang}
\affiliation{%
  \institution{Xianyu of Alibaba}
  \city{Hangzhou}
  \country{China}
}
\email{zhanglongtao.zlt@taobao.com}
\authornotemark[1]

\author{Ailong He}
\affiliation{%
  \institution{Xianyu of Alibaba}
  \city{Hangzhou}
  \country{China}
}
\email{along.hal@taobao.com}
\authornotemark[1]

\author{Shuzhi Cao}
\affiliation{%
  \institution{Xianyu of Alibaba}
  \city{Hangzhou}
  \country{China}
}
\email{caoshuzhi.csz@alibaba-inc.com}

\author{Shuguang Han}
\authornote{Corresponding Author}
\affiliation{%
  \institution{Xianyu of Alibaba}
  \city{Hangzhou}
  \country{China}
}
\email{shuguang.sh@taobao.com}

\author{Jufeng Chen}
\affiliation{%
  \institution{Xianyu of Alibaba}
  \city{Hangzhou}
  \country{China}
}
\email{jufeng.cjf@taobao.com}

\renewcommand{\shortauthors}{Zhihao Lv et al.}

\begin{abstract}
 Recommendation systems are essential for personalizing e-commerce shopping experiences. Among these, Trigger-Induced Recommendation (TIR) has emerged as a key scenario, which utilizes a trigger item—explicitly represents a user's instantaneous interest—to enable precise, real-time recommendations. Although several trigger-based techniques have been proposed, most of them struggle to address the intent myopia issue, that is, a recommendation system overemphasizes the role of trigger items and narrowly focuses on suggesting commodities that are highly relevant to trigger items. Meanwhile, existing methods rely on collaborative behavior patterns between trigger and recommended items to identify the user's preferences, yet the sparsity of ID-based interaction restricts their effectiveness. To this end, we propose the Deep Adaptive Intent-Aware Network (\textbf{DAIAN}) that dynamically adapts to users' intent preferences. In general, we first extract the users' personalized intent representations by analyzing the correlation between a user's click and the trigger item, and accordingly retrieve the user's related historical behaviors to mine the user’s diverse intent. Besides, sparse collaborative behaviors constrain the performance in capturing items associated with user intent. Hence, we reinforce similarity by leveraging a hybrid enhancer with ID and semantic information, followed by adaptive selection based on varying intents. Experimental results on public datasets and our industrial e-commerce datasets demonstrate the effectiveness of DAIAN.
\end{abstract}




\maketitle

\section{Introduction}
\label{sec 1}
\begin{figure}[htbp]
  \centering
  \includegraphics[width=\linewidth]{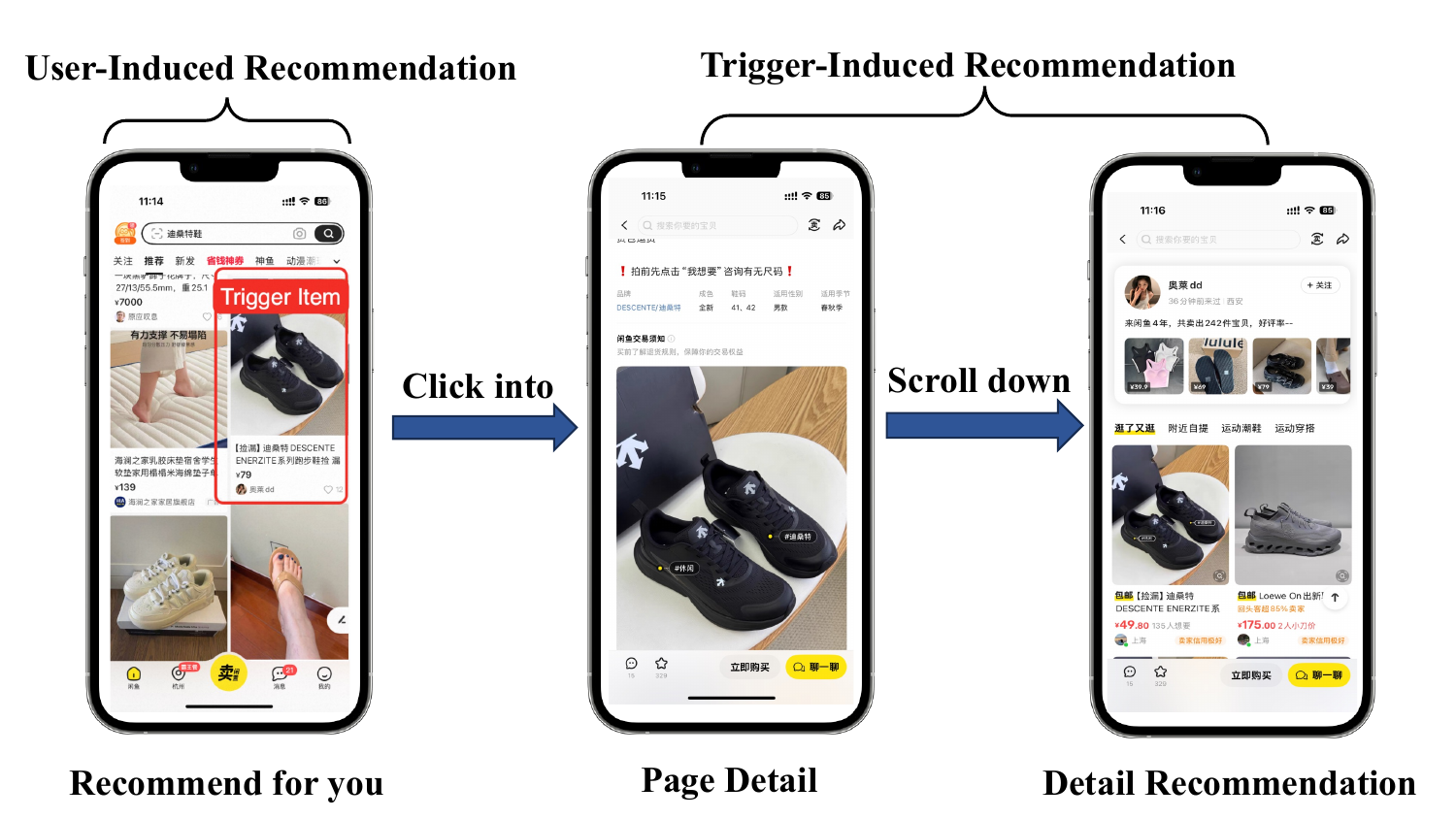}
  \caption{\textbf{Two primary recommendation scenarios on the Xianyu platform. The photo highlighted by a red solid frame represents User-Induced Recommendation. When users click on a trigger item and scroll down, they can transition to Trigger-Induced Recommendation.}}
  \label{intro}
  \vspace{-10pt}
\end{figure}

Recommendation systems have become indispensable in e-commerce platforms, improving user shopping experiences and increasing product sales. As China's largest consumer-to-consumer (C2C) e-commerce platform, Xianyu provides users with two key recommendation scenarios: User-Induced Recommendation (UIR) and Trigger-Induced Recommendation (TIR). As illustrated in the left part of Figure \ref{intro}, the {\itshape Recommend for You} module focuses on recommending items based on historical behavior. These similar scenarios generate items primarily based on user historical preferences and are known as UIR. Once
users click on an item, they are redirected to a specific scenario, e.g., {\itshape Page Detail }(the middle part of Figure \ref{intro}). As users scroll down, they can browse additional recommended products, most of which are contextually related to the initially clicked item. These scenarios are referred to as TIR and have evolved into a vital element for online digital commerce.

Traditional CTR (Click-Through Rate) prediction methods \cite{liu2019feature,sun2019bert4rec,xu2019recurrent} primarily focus on modeling user historical behavior to infer interest distributions.
These approaches achieve optimal performance in UIR scenarios within industrial environments. However, conventional paradigms often neglect the role of trigger items, leading to suboptimal performance when directly applied to TIR scenarios. Recently, TIR has attracted significant industry attention, prompting the development of novel methods that address TIR-specific challenges and show superior effectiveness compared to traditional approaches. 

\begin{figure}[htbp]
    \centering
    \begin{subfigure}[c]{0.48\textwidth}
        \centering
        \includegraphics[width=\linewidth]{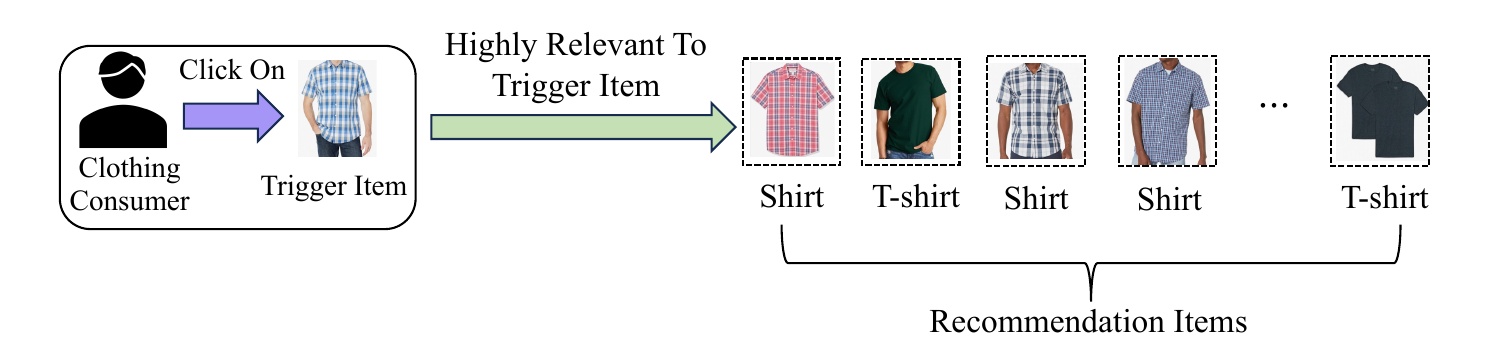}
        \caption{Existing TIR Solutions}
        \label{fig:sub1}
    \end{subfigure}
    \begin{subfigure}[c]{0.48\textwidth}
        \centering
        \includegraphics[width=\linewidth]{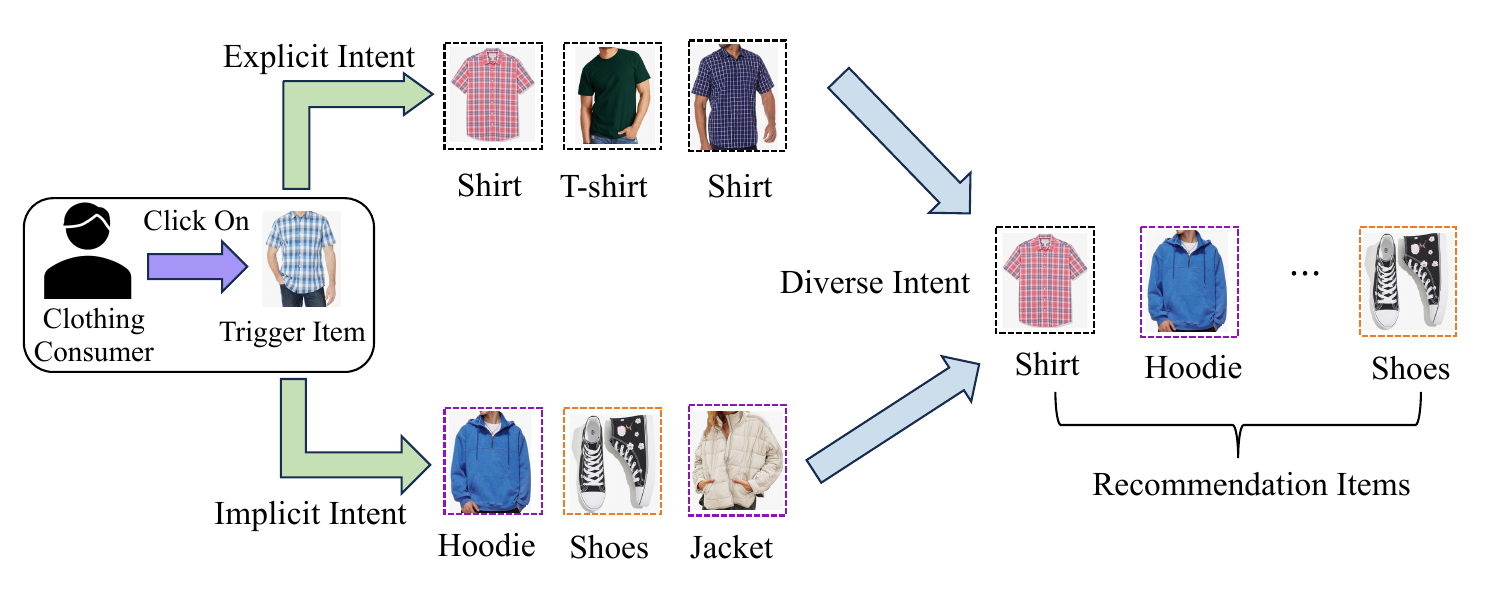}
        \caption{DAIAN}
        \vspace{-10pt}
        \label{fig:sub2}
    \end{subfigure}
    
    \caption{Differences in user intent preference modeling between existing TIR solutions and DAIAN. Black dashed boxes denote strongly-relevant items, purple dashed boxes represent relevant items, and orange dashed boxes highlight irrelevant items.}
    \label{compare}
    \vspace{-15pt}
\end{figure}

However, user behavior is inherently uncertain during the shopping process. Consumers show interest not only in similar items but also in complementary items. For instance, after clicking on a shirt, 43.5\% of users subsequently purchase strongly-relevant items (e.g., another shirt), 32.1\% rate relevant products (e.g., hoodie), and the remaining 24.4\% even click irrelevant items (e.g., shoes). Nevertheless, current approaches struggle to handle the problem of intent myopia, as they overly rely on the trigger item to infer user preferences. This drawback results in the formation of a filter bubble, where recommendation diversity decreases and users are confined to homogeneous content.
Figure \ref{fig:sub1} demonstrates that existing TIR solutions \cite{shen2022deep,xiao2024deep,ma2024modeling,gao2024collaborative,xie2021real} limit user intent to a constrained scope, leading to recommendation items that are only highly relevant to the trigger item. Therefore, 
as shown in Figure \ref{fig:sub2}, when a user clicks on the trigger item, we first divide user intent into two aspects: explicit intent and implicit intent. We then integrate items associated with each intent and recommend a diverse set of items to the user, satisfying their multiple purchasing needs.

After extracting users' diverse intents, it is crucial to recommend items based on users' intents. For example, explicit intents require strongly-relevant items with respect to the trigger item, while latent intents require a different strategy. Nonetheless, existing methods rely on collaborative item ID patterns to identify the similarity between items, the sparsity of interactions limits their performance. Moreover, directly optimizing the CTR binary classification loss via end-to-end training manner can hinder the model's convergence when combining the new module with the base module.

To address the aforementioned challenges, we propose a novel method called Deep Adaptive Intent-Aware Network (\textbf{DAIAN}), which dynamically adapts to users' intent preferences. Generally, we first model user intent with respect to the trigger item as a preference distribution by analyzing the click probabilities of items at multiple similarity levels to the trigger item. 
Specifically, we extract explicit intent from subsequences related to the trigger item and leverage the extracted intent distribution to explore latent user intent, thereby mining users' diverse intents. Additionally, semantic information and collaborative signals jointly strengthen similarity, and the shifting unit performs tailored selection for user' diverse intents to discover their corresponding preferences, thus addressing the issue of insufficient interaction. In addition, a three-stage training strategy is introduced to tackle the problem of in-convergence in the fusion model.

The significant contributions of this paper are as follows:
\begin{itemize}
\item We propose a novel model architecture named the Deep Adaptive Intent-Aware Network (DAIAN). Its primary objective is to address the challenge of intent myopia, a phenomenon where recommendation systems overemphasize trigger items at the expense of user intent diversity.
\item We first extract users' intent representations by analyzing the correlation between their clicks and the trigger item, and accordingly retrieve historical behaviors to mine their diverse intents. Additionally, we enhance similarity by leveraging a hybrid enhancer with ID and semantic information, and subsequently perform adaptive selection based on users' varying intents, to overcome the issue of sparse interaction in identifying users' interests.
\item Promising experiments on offline datasets and online industrial e-commerce platforms strongly demonstrate that our proposed model outperforms state-of-the-art baselines.
\end{itemize}

\section{Related Work}
\textbf{User Behavior Modeling.} Current user behavior modeling \cite{gharibshah2020deep,pi2020search,song2021coarse,zhou2018deep} primarily employs deep neural networks to capture representations of aggregated user interests from user behavior sequences. For instance, DIN \cite{zhou2018deep} proposes an attention mechanism that computes the relevance between historical behaviors and the target item. 
DIEN \cite{zhou2019deep} utilizes an interest evolution mechanism that integrates GRU with an attention mechanism to model the dynamic evolution of user interests. DSIN \cite{feng2019deep} segments the user behavior sequence into multiple sessions and uses a self-attention mechanism to extract users’ interests in each session. Although the aforementioned methods have demonstrated their effectiveness in various scenarios, the lack of consideration for modeling user instantaneous intent leads to suboptimal performance when applying them to TIR scenarios.

\textbf{Trigger-Induced Recommendation.} Driven by the limited effectiveness of conventional approaches in TIR scenarios, both industry and academic researchers have explored various improved solutions. DIHN \cite{shen2022deep} is the first work to study TIR problems, generating a precise probability score to predict user intent on the trigger item. DIAN \cite{xia2023deep} proposes a framework that utilizes conditional probability to balance the results of trigger-free and trigger-based recommendations based on the estimated intent. DEI2N \cite{xiao2024deep} considers the dynamic change of user immediate interest, temporal information, and interactions between trigger and target items to further improve CTR performance. Although the above existing trigger-based approaches achieve better results compared to traditional methods in TIR scenarios, they fail to address the issue of intent myopia, which causes challenges in user intent modeling and CTR prediction.

\textbf{Cold-start Recommendation.} Related research on cold-start problems focuses on representation generalization. POSO \cite{dai2021poso} enhances the model's ability to represent and generalize new items by introducing customized neural networks that are adapted to their specific features. MSNet \cite{wu2024metasplit} and MWUF \cite{zhu2021learning} leverage additional information from items and users to warm up the embedding of cold item IDs through a scaling and shifting network.

\section{Preliminaries}
In this section, we formulate the problem and provide a brief overview of the existing production CTR prediction model used in large-scale recommendation systems. 

\begin{figure*}[tbp]
  \centering
\includegraphics[width=0.9\linewidth,keepaspectratio]{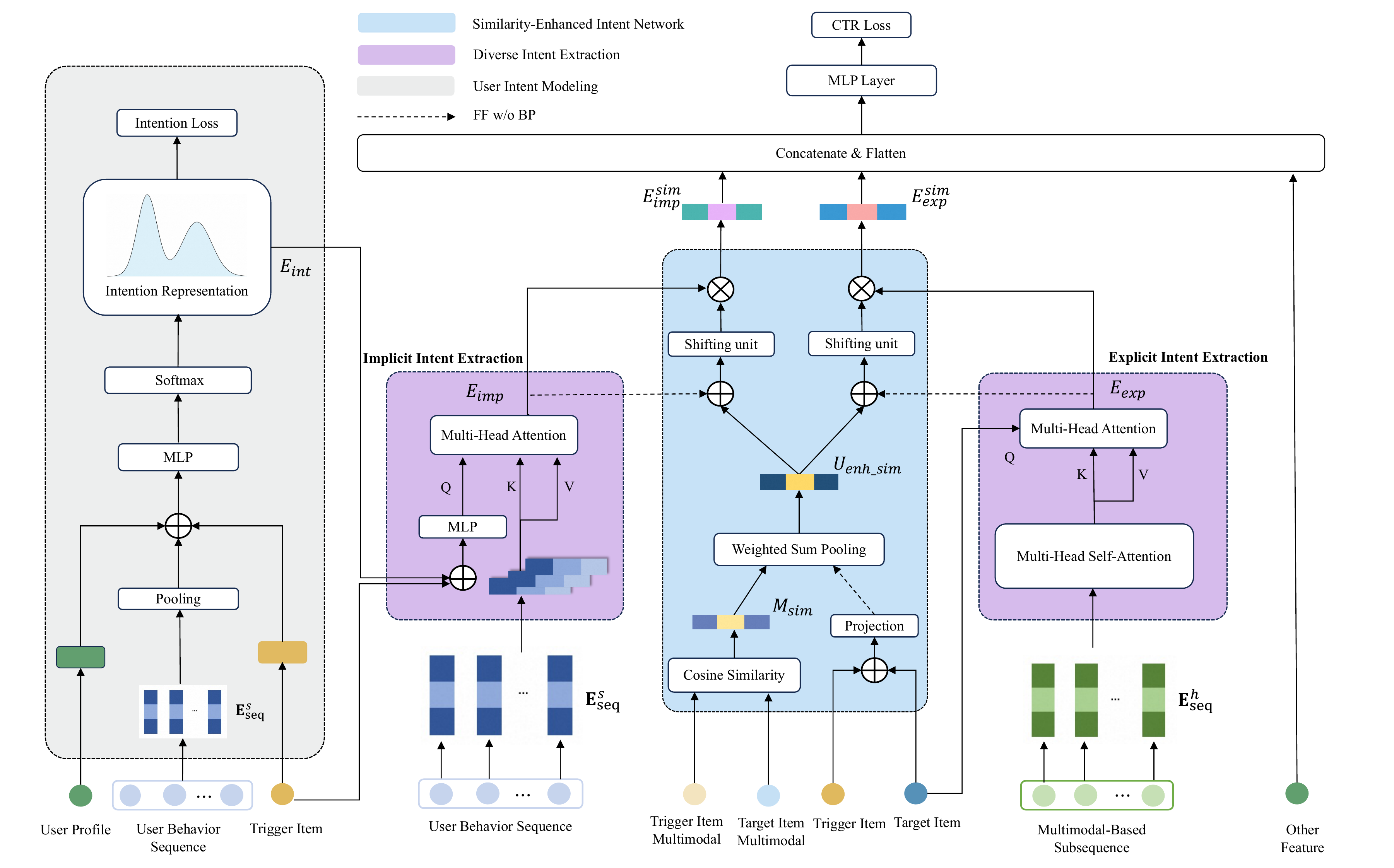}
  \caption{\textbf{An overall architecture of DAIAN can be broadly divided into three modules: (1) User Intent Modeling Module generates users’ personalized intent representations. (2) Diverse Intent Extraction Module retrieves the user’s historical behaviors to extract users' diverse intents. (3) Similarity-Enhanced Intent Network reinforces similarity and performs adaptive selection based on users' diverse intents.}}
  \label{arc}
  \vspace{-10pt}
\end{figure*}

\subsection{Problem Formulation}
Given a dataset 
\begin{math}
\mathcal{D} = \{\textbf{x},y\}^{N}, \{\textbf{x},y\}
\end{math}
marks a sample and {\itshape N} is the number of samples, where \textbf{x} denotes the high-dimensional feature vector composed of multi-fields (e.g., user and item fields), and 
\begin{math}
y
\end{math}
is the binary label, with 
\begin{math}
y = 1
\end{math}
indicating the sample is clicked. Our task is to accurately predict CTR probability
\begin{math}
p_{ctr} = p(y = 1|\textbf{x})
\end{math}
for the testing sample 
\begin{math}
    \textbf{x}
\end{math}
. Furthermore, it is essential to consider the influence of user instantaneous intent on CTR prediction tasks in TIR scenarios. The user's immediate intent prediction task aims to measure the personalized degree of users' intent with respect to the trigger item. Therefore, user intent can be represented as a probability distribution
\begin{math}
y_{int}
\end{math}
.

\subsection{Production CTR Prediction Model}
\label{sec 3.1}
We adopt Deep Interest Network \cite{zhou2018deep} as our base model due to its online efficiency and effectiveness. The model is structured with embedding layer and multi-layer perceptrons (MLPs) architecture, and it commonly integrates historical behavior modeling \cite{chen2022efficient,chang2023twin}.

\textbf{Embedding Layer}. The inputs are composed of non-sequential features (e.g., item IDs) and sequential features(e.g., user's historical behavior). The embedding layer maps each discrete feature in the raw input to a lower-dimensional vector. Non-sequential feature embeddings are simply concatenated, whereas sequential feature embeddings are processed by a sequence information extraction module to form a unified vector representation.
\textbf{Sequence Information Extraction}. Deep Interest Network (DIN) proposes an attention mechanism that computes the relevance between historical behavior and the target item. As a result, all relevant user behaviors are activated to calculate the final Click-Through Rate, leading ‌to better prediction performance. 

\section{Methodology}
The general architecture and detailed structure of our proposed DAIAN method are presented in Figure \ref{arc}, including three primary modules: User Intent Modeling (UIM), Diverse Intent Extraction (DIE), and Similarity-Enhanced Intent Network (SEIN). Additionally, we propose a novel Three-Stage Training Strategy to address the problem of non-convergence in fusing CTR prediction and intent modeling. The subsequent sections will elaborate on these modules and methods in detail.
\vspace{-5pt}

\subsection{User Intent Modeling}
\label{sec 4.1}
We first introduce the method to identify user intent for the trigger item in the sample dataset, which serves as the foundation for subsequent analysis.

\textbf{User Intent Labeling}. 
Several TIR methods omit user intent labels \cite{xiao2024deep,ma2024modeling}, resulting in a lack of supervision signals, which may introduce bias in the users' intent prediction. In addition, other approaches heavily rely on leaf categories \cite{shen2022deep,xia2023deep}, resulting in an overly narrow understanding of user intent that restricts the exploration of users' diverse interests.

Motivated by the recommendation systems with multimodal representation \cite{sheng2024enhancing,chen2016deep,elsayed2022end,yuan2023go,ge2018image,lynch2016images}, we utilize pre-trained representation obtained from the publicly available Qwen-VL model \cite{bai2023qwen,yang2025qwen3,chu2023qwen} to address the challenge of data homogeneity. This approach facilitates a more precise discernment of the user's intent for the trigger item. After clicking on a toothbrush, users may not only purchase strongly-relevant items (e.g., another toothbrush), but also buy relevant products (e.g., toothpaste), and even rate irrelevant products (e.g., clothes). Therefore, we divide the similarity between the trigger item and the target item into
\begin{math}
    n
\end{math}
different levels (i.e., Equal-Width Binning), each representing user intent for items at a constant similarity level. It can be formulated as a
\begin{math}
    n
\end{math}
-dimensional vector, which captures the user's preference distribution over items with varying degrees of similarity to the trigger item and is referred to as the ground-truth user intent distribution, denoted by
\begin{math}
y_{int}    
\end{math}
.
\begin{equation}
    y_{int} = \lbrack y_{1}, y_{2}, \cdots, y_{n} \rbrack,
\end{equation}
where 
\begin{math}
    y_{i}
\end{math}
represents the user's intent probability for products at the i-th similarity score level. These probabilities form a distribution, satisfying
\begin{math}
    y_{i} \in \lbrack0,1\rbrack
\end{math}
. Here
\begin{math}
    n
\end{math}
is a hyper-parameter
and represents the number of similarity score levels.

Taking 
\begin{math}
    y_{j}
\end{math}
as an example, 
\begin{math}
    y_{j} = \frac{N_{j}}{N}
\end{math},
when the user clicks a total of
\begin{math}
    N
\end{math}
products within the current request,
\begin{math}
    N_{j}
\end{math}
denotes the number of clicks corresponding to the j-th similarity level
. In addition, the degree of relevance between items is measured by the cosine similarity of their multimodal embedding. It should be noted that the posterior labeling strategy is infeasible for online serving because recommendations must be generated instantaneously upon the user entering the TIR. Therefore, we need to estimate user intent.

\textbf{User Intent Estimation}.
UIM takes three parts as the network input: 1) user profile features, such as user\_id, gender, city, etc. 2) User Behavior Sequence. 3) Trigger item features, such as item\_id, brand, price, etc. The above features are denoted as
\begin{math}
    E_{user},
E^{s}_{seq}
\end{math}
and
\begin{math}
    E_{trigger}
\end{math}
.
Specifically, the embedding representation of behavior sequences and a single item can be formulated as:
\begin{equation}
    E^{s}_{seq} =\{E^{s}_{1},
E^{s}_{2},
\ldots,
E^{s}_{m}
\} \in  \mathbb{R} ^{m \times d},
\end{equation}
\begin{equation}
    E^{s}_{i} = E(\mathcal{F}_{ID})
\oplus
E(\mathcal{F}_{Side}),
\end{equation}
where
\begin{math}
\oplus
\end{math}
denotes the concatenation operator and
\begin{math}
m
\end{math}
is the length of behavior sequences.
\begin{math}
E^{s}_{i},E(\mathcal{F}_{ID})
\end{math}
and
\begin{math}
    E(\mathcal{F}_{Side})
\end{math}
mean the item embedding, id embedding and side info embedding respectively. 

Considering that user intents for items at different similarity levels generally exhibit the distribution with multiple peaks and troughs. Therefore, as is illustrated in Figure \ref{arc}, we model the user diverse intent as a probability distribution via the softmax function, rather than as a single point estimate. Specifically, the predicted user's intent representation 
\begin{math}
    E_{int}
\end{math}
is an n-dimensional vector that is evaluated against the ground-truth user intent
\begin{math}
    y_{int}
\end{math}.
\begin{align}
E_{int} &= \lbrack \hat e_{1}, \hat e_{2}, \cdots, \hat e_{n} \rbrack \\
&= Softmax(MLP(    E_{user}\oplus
E^{s}_{seq}
\oplus E_{trigger}
)),
\end{align}

where 
\begin{math}
    E_{int} \in \mathbb{R}^{1 \times n}
\end{math}
, and 
\begin{math}
    \hat e_{i} \in \lbrack 0,1\rbrack
\end{math}
represent the estimated intent values at 
\begin{math}
    i
\end{math}
-th similarity level.

\subsection{Diverse Intent Extraction}
Following the acquisition of the user's intent distribution, we decompose the process of extracting diverse intent into implicit intent exploration and explicit intent exploitation, primarily utilizing the user’s historical behavior and personalized intent representation.

\textbf{Implicit Intent Exploration}. 
As discussed in section \ref{sec 1}, users exhibit implicit intent following a trigger item click, for example, purchasing strongly-relevant items, rating relevant products or even clicking irrelevant items. Nevertheless, previous methods \cite{shen2022deep,xia2023deep,xiao2024deep,ma2024modeling} lack the consideration of modeling the diversity of user intent and neglect the importance of exploring user implicit intent, leading to suboptimal performance in TIR scenarios. To address the above issue, we activate the interest relationship within the historical behavior sequence, which is applicable to both the intent representation and the target item.

Specifically, we design the Implicit Intent Exploration (I2E) to capture implicit user interests through historical behavior and modeled intent distribution  
\begin{math}
    E_{int}
\end{math}
matching. The inputs of I2E are composed of three parts: 1) user behavior sequence, denoted as 
\begin{math}
E^{s}_{seq} =\{E^{s}_{1},
E^{s}_{2},
\ldots,
E^{s}_{m},
\} \in  \mathbb{R} ^{m \times d}
\end{math} 
. 2) user's personalized intent representation
\begin{math}
E_{int}
\end{math} 
, which is generated by the User Intent Modeling Module. 3) trigger item,  expressed as
\begin{math}
E_{trigger}    
\end{math}
.

We first integrate the trigger item 
\begin{math}
    E_{trigger}
\end{math}
with personalized intent representation
\begin{math}
E_{int}    
\end{math}
to generate intent-based embedding 
\begin{math}
U_{int}    
\end{math}
, which is designed to better capture user implicit intent. Furthermore, we utilize Multi-Head Target Attention (MHTA) \cite{vaswani2017attention} to retrieve ambiguous intents within user behavior sequences
\begin{math}
E^{s}_{seq}    
\end{math}
. Finally, the output of this module is formulated as:
\vspace{-5pt}
\begin{equation}
    U_{int} = MLP(E_{int}\oplus E_{trigger}),
\end{equation}
\begin{equation}
\label{mhta}
    head_{i} = softmax(\frac
{W_{i}^{Q}U_{int} \cdot  (W_{i}^{K}E_{seq}^{s})^{\top}}
{\sqrt{d}}
)
W_{i}^{V}E_{seq}^{s},
\end{equation}
\begin{equation}
    E_{imp} = \lbrack head_{1}\oplus head_{2}\ldots\oplus head_{m}\rbrack
W^{O},
\end{equation}
where 
\begin{math}
    W_{i}^{Q},W_{i}^{K},W_{i}^{V} \in \mathbb{R}^{d\times d}
\end{math}
are projection matrices of the
\begin{math}
    i
\end{math}
-th head for the query, key and value respectively. The
number of heads 
is
\begin{math}
    m
\end{math}
and
\begin{math}
    W^{O} \in \mathbb{R}^{d\times d}
\end{math}
is the linear transformation matrix. The 
\begin{math}
    d
\end{math}
represents the dimension of each head.

\textbf{Explicit Intent Exploitation}. User behaviors exhibit a causal relationship and convey explicit preferences in TIR scenarios, where trigger items reflecting users' immediate and clear intent. For instance, when a user clicks on a trigger item belonging to the shirt category, it indicates that the user has a strong interest in items associated with the shirt category at this time. However, manually defined categories \cite{ma2024modeling,xiao2024deep,xia2023deep,shen2022deep} fail to capture textual and visual correlations within items, leading to limited matching accuracy. 

To address the above challenges, we propose Explicit Intent Extraction, which leverages multimodal information to replace the category-based selection of subsequence. The subsequence is aggregated to enhance the user's clear interest. 
As mentioned in Section \ref{sec 4.1}, similarity is determined by the cosine distance of embedding vectors, and we set thresholds for different similarity levels. The representation of the multimodal-based subsequence can be formulated as 
\begin{math}
        E^{h}_{seq} =\{E^{h}_{1},
E^{h}_{2},
\ldots,
E^{h}_{m}
\} \in  \mathbb{R} ^{m \times d}
\end{math}.

As shown in Figure 
\begin{math}
  \ref{arc}  
\end{math}
, we follow the same approach as in Eq.\ref{mhta} , but substitute intent-based item representation 
\begin{math}
    U_{int} 
\end{math}
with the target item embedding
\begin{math}
    \textbf{E}_{target} 
\end{math}
and replace user's historical sequence
\begin{math}
    E^{s}_{seq}
\end{math}
with subsequence
\begin{math}
    E^{h}_{seq}
\end{math}
. Notably, we apply Multi-Head Self-Attention (MHSA)  \cite{vaswani2017attention} to effectively extract the dependencies between representation pairs before feeding them into the attention unit. The output is given below.
\begin{equation}
        E_{seq}^{mhsa} = MHSA(E_{seq}^{h},E_{seq}^{h},E_{seq}^{h}),
\end{equation}
\begin{equation}
    E_{exp} = MHTA(E_{target}, E_{seq}^{mhsa}, E_{seq}^{mhsa}),
\end{equation}

\subsection{Similarity-Enhanced Intent Network}
Following the extraction of users' diverse intents, it is crucial to identify user's interest and precisely capture items associated with user intent. Therefore, we strengthen similarity through hybrid enhancer and perform adaptive interest selection via shifting units.

\textbf{Hybrid Similarity Enhancer}. 
It is widely recognized that online recommendation \cite{xie2021causcf, jin2012bayesian, wang2019neural} relies on ID-based collaborative signals to identify correlation between items. However, collaborative behavior frequency between trigger and recommended items is particularly sparse in TIR scenarios, leading to constrain their effectiveness. To reinforce the correlation between trigger items and target items, we combine multimodal features with item ID. Hybrid Similarity Enhancer takes two inputs: 1) target/trigger item ID, which is noted as 
\begin{math}
    E_{target},E_{trigger}
\end{math}
. 2) pre-trained embeddings of the target/trigger item are denoted them as 
\begin{math}
    M_{tar},M_{tri}
\end{math}
. Considering that collaborative features and multimodal information inherently exhibit feature-space misalignment \cite{liao2024llara}, we project collaborative ID embeddings
into the multimodal space through a trainable projection 
\begin{math}
    Proj
\end{math}
(i.e., multi-layer MLP). 
\begin{equation}
    U_{co\_id} = Proj(E_{target} \oplus E_{trigger} ),
\end{equation}

\begin{equation}
    M_{sim} = \text{Cosine}(M_{tar}, M_{tri}) = \frac{M_{tar} \cdot M_{tri}}{\|M_{tar}\| \cdot \|M_{tri}\|},
\end{equation}

where 
\begin{math}
    M_{sim}, U_{co\_id}
\end{math}
represents the semantic similarity, collaborative ID embedding respectively.

Subsequently, we bin the semantic similarity 
\begin{math}
    M_{sim}
\end{math}
, which can convert a scalar value into a multi-dimensional feature representation 
\begin{math}
    U_{sim}
\end{math}
. Moreover, we want to make more use of semantic similarity when the collaborative ID is relatively weak, and minimize its use when the collaborative ID is relatively strong. To this end, we integrate the semantic similarity embedding with the collaborative ID embedding to form the final representation
\begin{math}
    U_{enh\_sim}
\end{math}
by norm ratio.
\begin{equation}
    U_{enh\_sim} = \delta \cdot U_{sim} + (1-\delta) \cdot (\oslash (U_{co\_id})),
\end{equation}
\begin{equation}
    \delta = \frac{Norm(U_{sim})}{Norm(U_{sim}) + Norm( U_{co\_id})},
\end{equation}
where we prevent interference with the ID embedding updates by blocking gradient backpropagation, denoted as
\begin{math}
\oslash (\ast)    
\end{math}
.

\textbf{Intent-Aware Adaptive Network}.
In industrial-scale recommendation systems \cite{pi2020search,chang2023twin,chen2022efficient}, the dimension of the bottom-layer embedding is large, and directly feeding extracted interest representations into the embedding layer leads to them being overwhelmed, thereby limiting their effectiveness. Intent-Aware Adaptive Network (IA2N), introduces a adaptive shifting mechanism that incorporates intent as part of the input. This mechanism enables adaptive modification of weights based on the correlation between the intent and similarity, thereby preventing the representation from being neglected.

It takes three inputs: 1) implicit intent representation. 2) explicit intent representation. 3) enhanced similarity between trigger items and target items. The parts presented above are all derived from the preceding module's output, and we denote them as:
\begin{math}
    E_{imp},E_{exp}
\end{math}
and
\begin{math}
    U_{enh\_sim}
\end{math}
. Specifically, we concatenate enhanced similarity and intent embedding, then feed them into the shifting unit 
\begin{math}
    SU 
\end{math}
(i.e., Gate NU \cite{chang2023pepnet} and multi-layer MLP) to generate the weighted vector
\begin{math}
    \delta
\end{math}
.
\begin{equation}
    \delta_{exp} = SU(U_{enh\_sim}\oplus((\oslash E_{exp})),
\end{equation}
\begin{equation}
    \delta_{imp} = SU(U_{enh\_sim}\oplus((\oslash E_{imp})),
\end{equation}
where we avoid interference with the intent embedding updates by blocking gradient backpropagation, denoted as
\begin{math}
    \oslash (\ast)
\end{math}.

Subsequently, we apply the weighted vector 
\begin{math}
    \delta
\end{math}
to execute adaptive transformations on the embeddings 
\begin{math}
    E_{exp}
\end{math}
and 
\begin{math}
    E_{imp}
\end{math}
. The transformed embeddings are as below.
\begin{equation}
    E_{exp}^{sim} = \delta_{exp} \otimes E_{exp},
\end{equation}
\begin{equation}
    E_{imp}^{sim} = \delta_{imp} \otimes E_{imp},
\end{equation}

where 
\begin{math}
    E_{exp}^{sim}
\end{math}
and 
\begin{math}
    E_{imp}^{sim}    
\end{math}
are the explicit intent related and implicit intent related interest representations.

\subsection{Three-Stage Training Strategy}
\label{sec 4.4}
\begin{figure}[tbp]
  \centering
  \includegraphics[width=\linewidth]{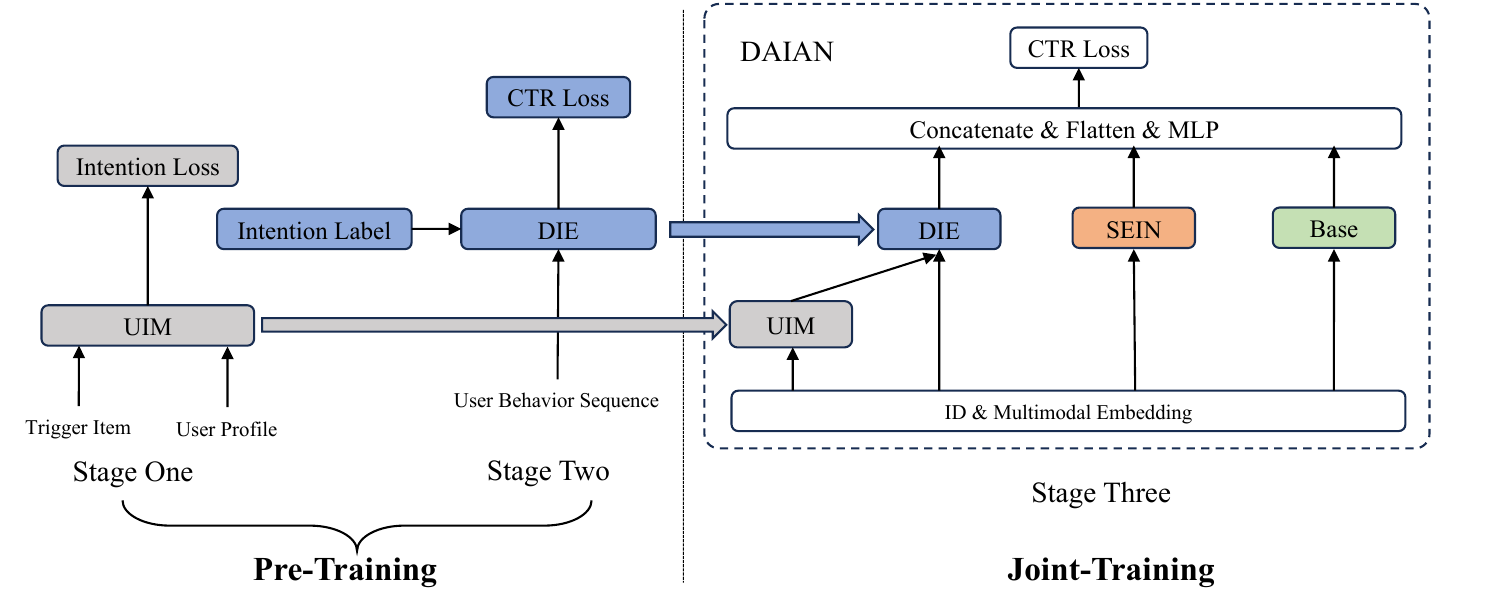}
  \caption{\textbf{An Overview of our Three-Stage Training Strategy: the pre-training of UIM and DIE, followed by the integration of pre-trained users' personalized intent and  diverse intent representations into recommendation model.}}
  \label{threestep}
  \vspace{-10pt}
\end{figure}

We have merged User Intent Modeling and CTR prediction. Nevertheless, directly optimizing the CTR prediction cross entropy loss via end-to-end training manner may hinder the model’s convergence. This is because the click label is a too-weak supervised signal for all the
components of the model to converge to their true physical meaning, and the end-to-end loss function makes the optimization process difficult. Therefore, to address the non-convergence of each module, we propose a three-stage training strategy, as shown in Figure \ref{threestep}. Specifically, utilizing the user's ground-truth intent 
\begin{math}
    y_{int}
\end{math}
as the supervised signal, we first suppose 
\begin{math}
    g(\textbf{x};\theta)
\end{math}
represents the UIM parameter by
\begin{math}
    \theta
\end{math} and pretrain the UIM to initialize its parameters. As discussed in section \ref{sec 4.1},
\begin{math}
    y_{int}
\end{math}
is a n-dimensional vector. 
To align the model's prediction 
\begin{math}
    E_{int}
\end{math}
with ground-truth intent
\begin{math}
    y_{int}
\end{math}
, we introduce a regularization term to reinforce their consistency by minimizing the KL divergence between the two distributions. The overall objective for the UIM pre-training task is expressed as:
\begin{equation}
    T_{pretrain}(\theta) = D_{KL}(y_{int}||E_{int}),
\end{equation}

Given that the input of DIE incorporates the estimated personalized intent
\begin{math}
    E_{int}
\end{math}
, we then pretrain the DIE parameter
\begin{math}
    \eta
\end{math}
through replacing the predicted value
\begin{math}
    E_{int}
\end{math}
with ground-truth intent
\begin{math}
    y_{int}
\end{math}
, thereby accelerating the convergence of parameter
\begin{math}
    \eta
\end{math}
. Additionally, we utilize the CTR prediction objective as a supervision signal.
\begin{equation}
\begin{split}
    T_{pretrain}(\eta) = -\frac{1}{|\mathcal{D}|} \sum_{(\textbf{x},y)\in\mathcal{D}}
(ylogf(\textbf{x},y_{int}) \\
+(1-y)\log(1-f(\textbf{x},y_{int})),
\end{split}
\end{equation}

Finally, we jointly train UIM, DIE, SEIN and the CTR base module with the primary objective of optimizing the CTR prediction task. Notably, due to the inconsistency in the optimization objectives between the CTR and user intent tasks, parameters are not shared between the corresponding networks. Moreover, we remove intent signal supervision and fine-tune the UIM with a smaller learning rate, as the main goal is to optimize the CTR prediction objective. Let
\begin{math}
    \varphi
\end{math}
denote the parameter of CTR base module. The cross-entropy loss of the CTR task is shown below:
\begin{align}
\begin{split}
    T_{ctr}(\theta,\eta,\varphi) = &-\frac{1}{|\mathcal{D}|} \sum_{(\textbf{x},y)\in\mathcal{D}}
(y\log\hat{y} +(1-y)\log(1-\hat{y})),
 \end{split}
 \label{final_loss}
\end{align}
where
\begin{math}
    \mathcal{D}
\end{math}
is the training set, and
\begin{math}
    y \in \{0,1\}
\end{math}
is the label for CTR prediction task.

\begin{table*}
  \caption{Offline Experiments Comparison of our method with competitors on two real-world datasets. The bold number in each column indicates the best result. The improvements are
statistically significant (one-sided rank-num \textmd{p-value < 0.01}) over compared methods.}
  \label{tab:offline}
  \renewcommand{\arraystretch}{1}
     \setlength{\tabcolsep}{5pt}  
\begin{tabular}{c|cccc|cccc}
\toprule
\multirow{2}{*}{\centering Model} & \multicolumn{4}{c|}{Xianyu} & \multicolumn{4}{c}{Alimama} \\
\cline{2-9}
                      & AUC & RelaImpr & GAUC & RelaImpr
                      & AUC & RelaImpr & GAUC & RelaImpr 
                      \\
\hline
DIN &$0.6914$ $\pm\ $ $0.0013$ &$0.00\%$ &$0.6318$ $\pm\ $ $0.0010$  &$0.00\%$
&$0.6154$ $\pm\ $ $0.0007$  &$0.00\%$ &$0.5954$ $\pm\ $ $0.0008$ &$0.00\%$                   \\
DIEN & $0.6915$ $\pm\ $ $0.0022$ & $0.05\%$ & $0.6321$ $\pm\ $ $0.0019$ & $0.23\%$
& $0.6155$ $\pm\ $ $0.0005$ & $0.09\%$ & $0.5957$ $\pm\ $ $0.0012$ & $0.31\%$                     \\
POSO & $0.6908$ $\pm\ $ $0.0012$ & $-0.31\%$ & $0.6315$ $\pm\ $ $0.0013$ & $-0.23\%$
& $0.6159$ $\pm\ $ $0.0009$ & $0.43\%$ & $0.5959$ $\pm\ $ $0.0006$ & $0.52\%$
\\
MWUF & $0.6920$ $\pm\ $ $0.0009$ & $0.31\%$  & $0.6341$ $\pm\ $ $0.0010$ & $1.75\%$
& $0.6162$ $\pm\ $ $0.0006$ & $0.69\%$ & $0.5963$ $\pm\ $ $0.0002$ & $0.94\%$
        \\
DIHN & $0.6947$ $\pm\ $ $0.0003$ & $1.72\%$  & $0.6347$ $\pm\ $ $0.0005$ & $2.20\%$
& $0.6166$ $\pm\ $ $0.0008$ & $1.04\%$ & $0.5966$ $\pm\ $ $0.0010$ & $1.26\%$                         \\
DIAN & $0.6962$ $\pm\ $ $0.0013$ & $2.51\%$ & $0.6362$ $\pm\ $ $0.0012$ & $3.34\%$
& $0.6176$ $\pm\ $ $0.0002$ & $1.91\%$ & $0.5974$ $\pm\ $ $0.0007$ & $2.10\%$                         \\
DEI2N & $0.6986$ $\pm\ $ $0.0011$ & $3.76\%$ & $0.6396$ $\pm\ $ $0.0010$ & $5.72\%$
& $0.6194$ $\pm\ $ $0.0003$ & $3.47\%$ & $0.5984$ $\pm\ $ $0.0005$ & $3.14\%$                      \\
\hline
DAIAN & $\boldsymbol{0.7024 \pm\ 0.0008}$ & $\boldsymbol{5.75\%}$
& $\boldsymbol{0.6432 \pm\ 0.0007}$ & $\boldsymbol{8.65\%}$
& $\boldsymbol{0.6259 \pm\ 0.0002}$ & $\boldsymbol{9.10\%}$
& $\boldsymbol{0.6062 \pm\ 0.0003}$ & $\boldsymbol{11.32\%}$
\\
\bottomrule
\end{tabular}
\end{table*}

\begin{table*}
  \caption{Ablation study of different DAIAN variants on Xianyu dataset, where "w/o" is short for "without".}
    \renewcommand{\arraystretch}{1}
     \setlength{\tabcolsep}{17pt}  
  \label{tab:ablation}
\begin{tabular}{c|cc|cc}
\toprule
 Model & AUC & RelaImpr & GAUC & RelaImpr             \\
\hline
DAIAN w/o UIM &$0.6970$ $\pm\ $ $0.0011$ &$-2.67\%$          & $0.6376$ $\pm\ $ $0.0009$ & $-3.91\%$                \\
DAIAN w/o DIE & $0.7004$ $\pm\ $ $0.0009$ & $-0.99\%$          & $0.6404$ $\pm\ $ $0.0011$ & $-1.96\%$             \\
DAIAN w/o SEIN & $0.6984$ $\pm\ $ $0.0007$ & $-1.98\%$         & $0.6382$ $\pm\ $ $0.0012$ & $$-3.49\%$$                \\
DAIAN w/o Three-Stage Training & $0.6997$ $\pm\ $ $0.0012$ & $-1.33\%$  & $0.6396$ $\pm\ $ $0.0013$ & $-2.51\%$                      \\
\hline
DAIAN & $\boldsymbol{0.7024 \pm\ 0.0008}$ & $\boldsymbol{0.00\%}$ & $\boldsymbol{0.6432 \pm\ 0.0007}$ & $\boldsymbol{0.00\%}$                  \\
\bottomrule
\end{tabular}
\end{table*}

\section{Experiments}
In this section, we conduct comprehensive experiments on both the offline dataset and online A/B testing to evaluate the performance of our proposed method.

\subsection{Experimental Setup}
\textbf{Datasets}. Two real-world datasets are used in experiments. Dataset details are as follows.

\textbf{Xianyu dataset}. We collect traffic logs of 8 days from the Xianyu TIR scenario to build the industrial production dataset. The daily training dataset includes 1 billion records and each record contains 818 features (e.g., user, trigger item and target item features). We take 8 days of data for the experiment performance evaluation, in which data from the first 7 days is used for model training and the remaining data of 1 day is used for model testing.

\textbf{Alimama} \footnote{https://tianchi.aliyun.com/dataset/56}. It is a representative e-commerce dataset provided by Alimama advertising platform. As there are no logs for trigger items, we follow the data processing strategy by \cite{shen2022deep} and the trigger item is defined as users' most recent clicks within a 4-hour window. The daily training dataset includes 26 million records and each record contains 129 features (e.g., user, ad and item features). We take 8 days of data for the experiment performance evaluation, in which data from the first 7 days is used for training, and the remaining data of 1 day is used for testing.

\textbf{Compared Methods}. 
To demonstrate the superiority of our proposed method, we primarily compare it with the following baseline approaches.
\begin{itemize}

\item \textbf{DIN \cite{zhou2018deep}} proposes an attention mechanism that computes the relevance between historical behavior and the target item to aggregate historical behavior information. 

\item \textbf{DIEN \cite{zhou2019deep}} utilizes an interest evolution mechanism that integrates GRU with an attention mechanism to model the dynamic evolution of user interests. 

\item \textbf{MWUF \cite{zhu2021learning}} leverages additional information from items and users to warm up cold item ID embeddings
through a scaling and shifting network. 

\item \textbf{POSO \cite{dai2021poso}} enhances
the generalization and robustness of models for new items by
designing specialized networks tailored to their characteristics.

\item \textbf{DIHN \cite{shen2022deep}} is the first work to study TIR problems, which generates a precise probability score to predict user’s instantaneous intent on the trigger item.

\item \textbf{DIAN \cite{xia2023deep}} designs a framework that utilizes conditional probability to balance the results of trigger-free and trigger-based recommendations based on the estimated intent.

\item \textbf{DEI2N \cite{xiao2024deep}} considers the dynamic change of user's immediate interest, temporal information, and interactions between trigger and target items to further improve CTR performance in TIR scenarios.

\end{itemize}

\textbf{Evaluation Metric}. 
The main evaluation metrics used in the experiments are defined as follows:
\begin{itemize}[leftmargin=*]
\item \textbf{AUC:} The Area Under the Curve (AUC) is a standard assessment measure in CTR tasks, reflecting the model's capacity to distinguish between positive and negative items. It essentially calculates the probability that the model assigns a higher score to a randomly selected positive item compared to a negative one. Therefore, the AUC can be calculated as:
\begin{equation}
    \text{AUC} = \frac{1}{|P||N|}\sum_{p\in P}\sum_{n\in N}
I(\Theta(p) > \Theta(n)),
\end{equation}
where 
\begin{math}
    P
\end{math}
and
\begin{math}
    N
\end{math}
mean the positive and negative item set respectively.
\begin{math}
    \Theta
\end{math}
is the estimator function and 
\begin{math}
    I
\end{math}
is the indicator function.

\item \textbf{GAUC:} In contrast to AUC, which evaluates the model's overall ranking performance, the Group Area Under the Curve (GAUC) is intended to assess the quality of ordering based on the ranking of items within different user groups. Specifically, we first compute the AUC for each individual user and then take an average to obtain the final GAUC. Mathematically, it is defined as:
\begin{equation}
    \text{GAUC} = \frac{\sum_{i=1}^{n}(\#\text{impression}_{i}\times \text{AUC}_{i})}
    {\sum_{i=1}^{n}(\# \text{impression}_{i})}
\end{equation}
where 
\begin{math}
    \text{AUC}_{i}
\end{math}
stands for the AUC for 
\begin{math}
    i
\end{math}
-th user,
\begin{math}
    \#impression
\end{math}
 is its corresponding weight, and 
 \begin{math}
     n
 \end{math}
denotes the total number of users.

\item \textbf{RelaImpr:} RelaImpr metric is used to measure relative improvement over models and 0.5 stands for the AUC of a random guesser. Generally, higher values for AUC and GAUC indicate better performance. It can be formulated as:
\begin{equation}
    \text{RelaImpr} = (\frac{\text{AUC}(\text{measured model}) - 0.5}
    {\text{AUC}(\text{base model}) - 0.5}) \times 100\%
\end{equation}
\end{itemize}

\textbf{Implementation Details}. In all experimental setups, the batch size is set to 4096, and the AdaGrad optimizer is employed for parameter updates across all models, with the learning rate configured at 0.05. The DNN component of the above model architecture is identical, i.e., a three-layer MLP network with 512, 256, and 128 hidden units. The hidden units in each attention layer of the models are configured to 128, while the number of multi-head is set to 4. Besides, the user historical sequence is collected within last 14 days and the maximum length is 50.  


\subsection{Experimental Results}
In this section, we mainly discuss the following five aspects: offline experiments comparison, ablation study, online A/B testing, precision of user intent estimation and impact on different user group.

\textbf{(1) Offline Experiments Comparison}. 
To demonstrate the effectiveness and superiority of the DAIAN, we conduct offline experiments in the industrial environment using two real-world datasets: Xianyu and Alimama. The experimental results are compared with state-of-the-art (SOTA) approaches, which are presented in Table~\ref{tab:offline}. Our method achieved the best overall performance, outperforming previous approaches. Specifically, compared to the baseline model DIN, our proposed DAIAN achieved a 5.75\% AUC improvement and a 8.65\% GAUC increase on the Xianyu dataset. Additionally, on the Alimama dataset, our proposed DAIAN demonstrated superior performance, achieving an AUC of 0.6259 and a GAUC of 0.6062. Furthermore, our method outperforms traditional user behavior modeling paradigms, cold start approaches and existing methods, achieving the best AUC metrics.

\textbf{(2) Ablation Study}. 
To assess the contribution of the component module, an ablation study was conducted on our proposed DAIAN. The ablation experimental results on the Xianyu dataset are shown in Table~\ref{tab:ablation}. When users' intent is not modeled as a probability distribution, the GAUC drops from 0.6432 to 0.6376, demonstrating the importance of personalized intent extraction. In addition, without reinforcing similarity and executing adaptive selection for diverse intents, DAIAN w/o SEIN degrades the GAUC value from 0.6432 to 0.6382. It highlights the role of semantic similarity and shifting units in identifying users' interests. Meanwhile, we observe that the Diverse Intent Extraction and Three-Stage Training Strategy also enhance the model's performance.

\textbf{(3) Online A/B Testing}. 
We deploy the proposed DAIAN on Xianyu’s TIR scenario to evaluate its online performance. A long-term promising online experiment was conducted from June 1 to June 30, 2025, to evaluate DAIAN against the baseline model DIN, which accounted for 20\% of the total daily traffic. DAIAN achieves a 1.59\% increase in CTR, a 1.73\% increase in recommendation diversity and a 2.37\% rise in bills, demonstrating its superiority.

\begin{figure}[h]
	\centering
	\begin{subfigure}{0.495\linewidth}
		\centering
		\includegraphics[width=1\linewidth]{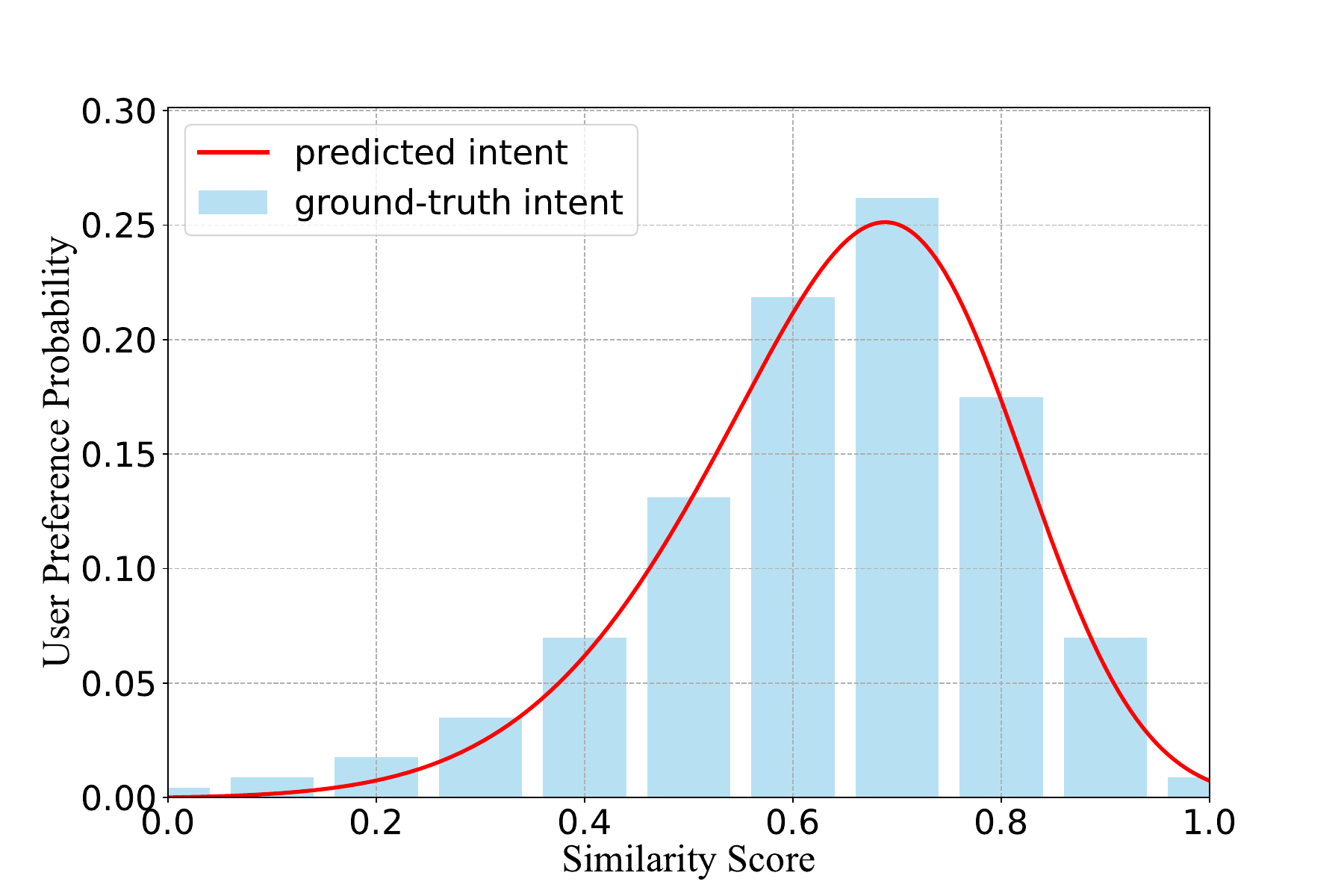}
		\caption{Specific-Intent Users}
		\label{intent_high}
        \vspace{-5pt}
	\end{subfigure}
	\centering
	\begin{subfigure}{0.495\linewidth}
		\centering
        \includegraphics[width=1\linewidth]{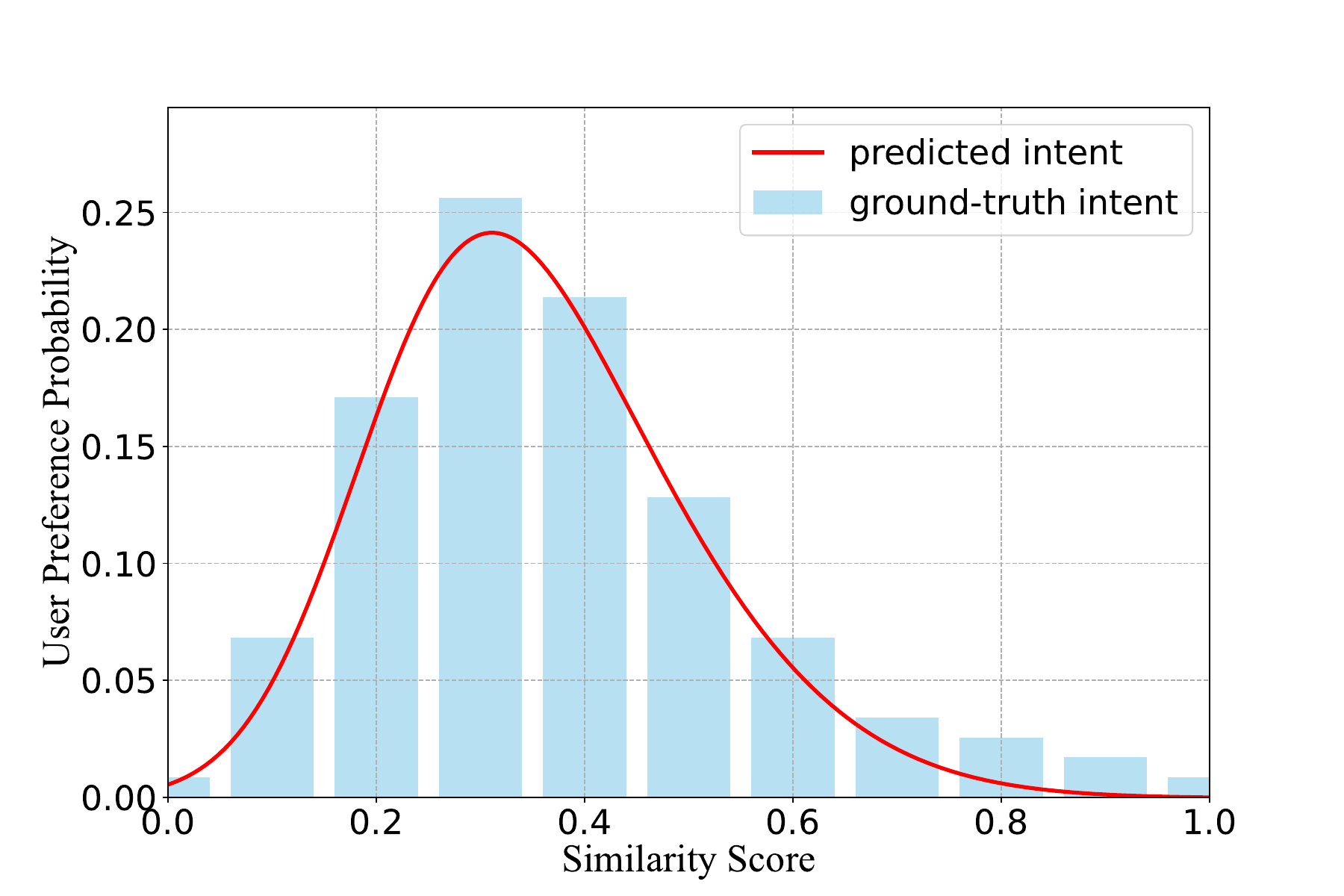}
        \caption{Broad-Intent Users}
		\label{intent_low}
        \vspace{-5pt}
	\end{subfigure}
	\caption{(a) exhibits the Predicted vs. Ground-Truth intent distribution for specific-intent users, and (b) exhibits the Predicted vs. Ground-Truth intent distribution for broad-intent users.}
	\label{fig1new}
    \vspace{-10pt}
\end{figure}

\textbf{(4) Precision of User Intent Estimation}.
To better assess the accuracy of intent representation 
\begin{math}
E_{int}    
\end{math}
, we firstly clarify its definition: 
\begin{math}
E_{int}    
\end{math}
models the user's conditional preference distribution over items with varying degrees of similarity to the trigger item, given that the user has clicked on it. We conduct large-scale sampling from the real-world data of two distinct user populations: Specific-Intent Users, who exhibit a high degree of interest in trigger items, and Broad-Intent Users, who display a lower level of interest. Subsequently, we measure the divergence between the predicted intent representation
\begin{math}
E_{int}    
\end{math}
and the ground-truth intent distribution
\begin{math}
    y_{int}
\end{math} both qualitatively, through visualization plots, and quantitatively, by calculating the JS divergence.
As depicted in Figure \ref{fig1new}, we visualize the comparison between
\begin{math}
y_{int}     
\end{math}
(represented by a bar plot) and 
\begin{math}
E_{int}   
\end{math}
 (illustrated by a line plot), thereby illustrating the goodness of fit between the two distributions. In this figure, the horizontal axis denotes the similarity score between the trigger and target items, and the vertical axis represents the corresponding user preference probability. Besides, A lower JS divergence value indicates a higher degree of consistency. Figure \ref{intent_high} reveals that for specific-intent users,
\begin{math}
E_{int}    
\end{math}
and 
\begin{math}
    y_{int}
\end{math} align closely, both indicating a strong preference for items identical to the trigger item, as indicated by a JS divergence of just 0.082. A similar pattern of high consistency is also observed for broad-intent users Figure \ref{intent_low}; however, their preferences are concentrated on items with low similarity to the trigger item, indicating a clear exploratory or divergent intent. The above finding demonstrates that despite the starkly contrasting preference distributions of the two populations, our proposed DAIAN remains feasible to achieve a precise fit for each, thereby enabling personalized modeling of user intent.
\vspace{-10pt}

\begin{figure}[htbp]
  \centering
  \includegraphics[width=0.8\linewidth]{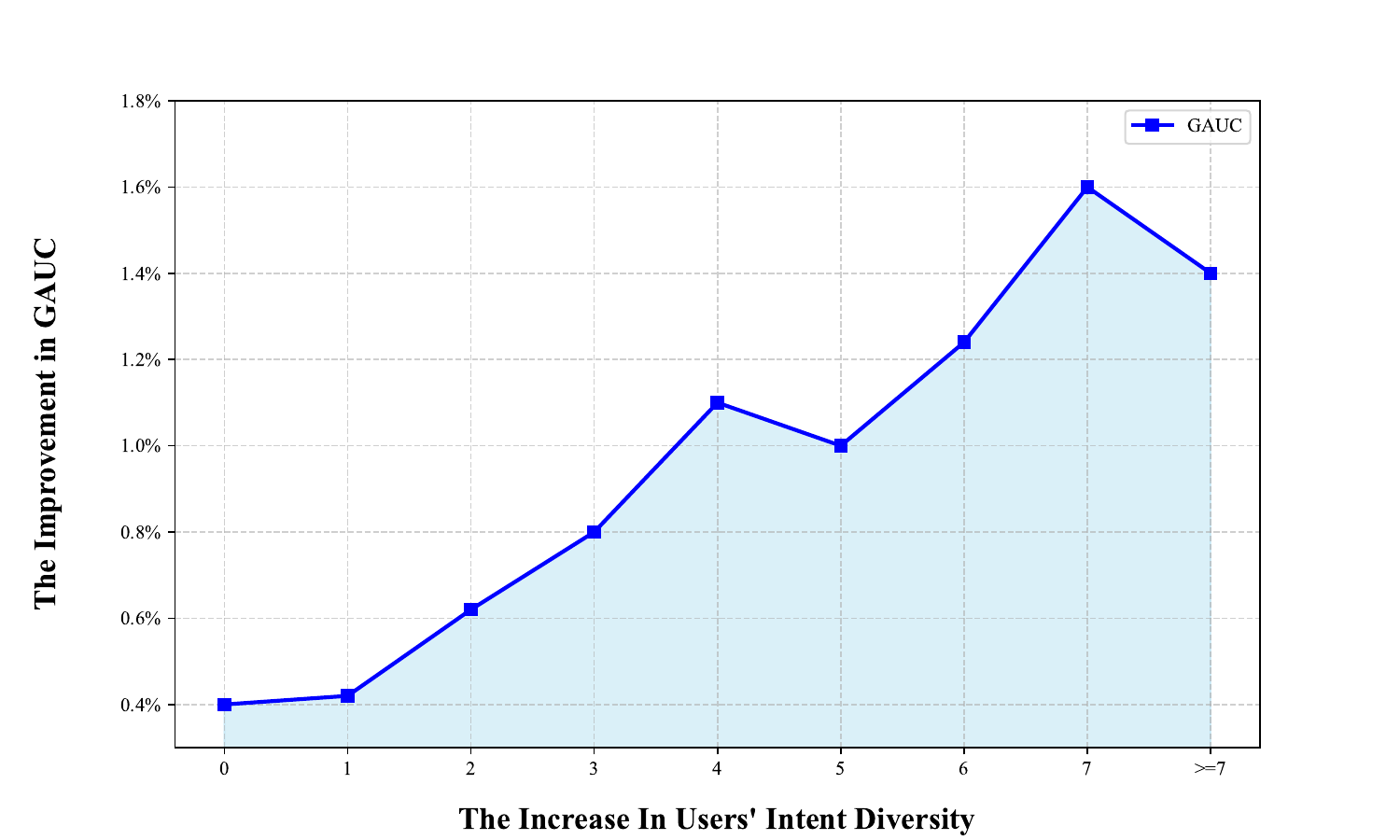}
  \caption{\textbf{The improvement in GAUC of different user groups split by the users' intent diversity.}}
  \label{product_analysis}
\end{figure}

\textbf{(5) Impact On Different User Group}.
To better evaluate the effectiveness of extracting users' diverse intents, we group users by the diversity of their intents, which is quantified as the number of leaf categories in their clicks per request that differs from the trigger item's leaf category. The diversity of users' intents is calculated in the last week of training data and we compute the GAUC metrics for each user group. The results in Figure \ref{product_analysis} demonstrate a strong positive correlation between the GAUC improvement and the increase in users' intent diversity. Notably, The GAUC of users with no intent diversity increased by only \textbf{0.4\%}; whereas with an intent diversity of 7 or more achieved a significant improvement of \textbf{1.6\%}. This finding further highlights that our proposed method DAIAN efficiently extracts user's diverse intents and satisfies user's personalized shopping needs.
\vspace{-10pt}

\vspace{-10pt}

\subsection{Hyperparameter Analysis}
\begin{figure}[htbp]
  \centering
  \includegraphics[width=0.76\linewidth]{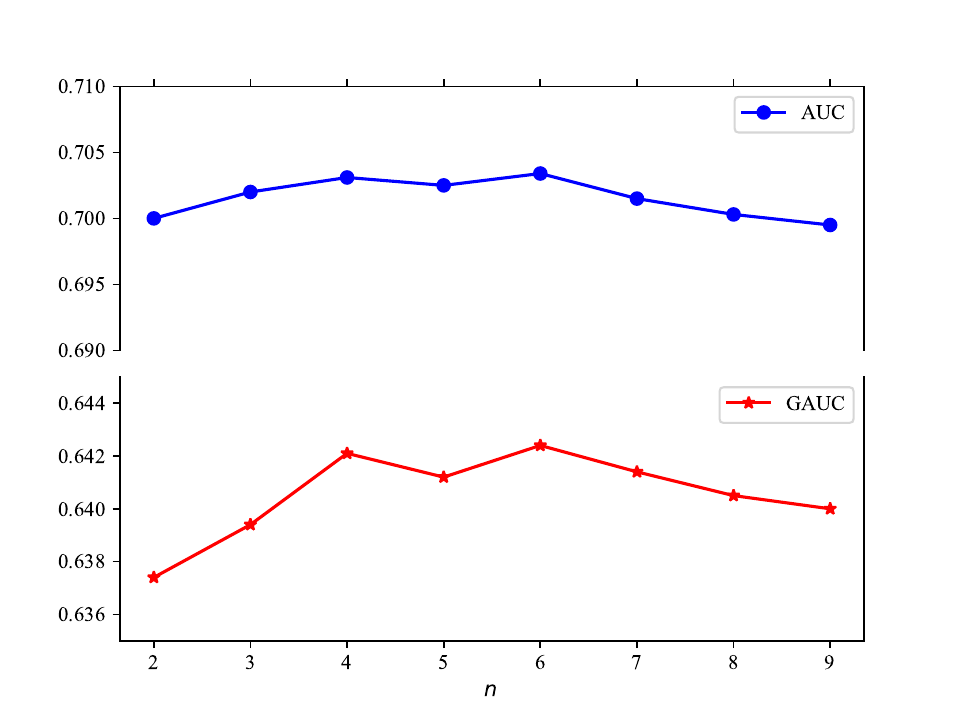}
  \caption{\textbf{The impact of the hyperparameter $n$ in the user intent labeling.}}
  \label{hyperpara}
  \vspace{-5pt}
\end{figure}

\textbf{The tuning of}
\begin{math}
    n
\end{math}
\textbf{in the user intent labeling.}
 As mentioned in section \ref{sec 4.1}, 
\begin{math}
 n 
\end{math}
is a hyper-parameter, which represents the number of similarity score level during User Intent Labeling and is utilized to analyze users' intent for target items that vary in similarity relative to trigger items. To determine the optimal value, we qualitatively explore the influence of different values in DAIAN, which is illustrated in Figure \ref{hyperpara}. It is obvious that when 
\begin{math}
    n
\end{math}
is set to 6, the model attains the best AUC and GAUC performance collectively.
\vspace{-9pt}

\vspace{-10pt}
\section{Conclusion}
In this paper, we designed a novel model structure, called Deep Adaptive Intent-Aware Network (\textbf{DAIAN}), to extract user's diverse intents and overcome the issue of sparse interaction in identifying users' interests in TIR scenarios. The proposed DAIAN consists of three key modules: 1) User Intent Modeling represents users' intent as a probability distribution and learns user preference patterns for items with different similarity to the trigger item. 2) Diverse Intent Extraction explores implicit intents and exploits explicit intents by retrieving the user’s related historical behaviors. 3) Similarity-Enhanced Intent Network aims to capture users' preferences by strengthening similarity and performing adaptive selection for varying intents. Additionally, comprehensive and confident experimental results have sufficiently validated the superiority of DAIAN over state-of-the-art baselines.


\bibliographystyle{ACM-Reference-Format}
\bibliography{sample-base}


\end{document}